\documentclass[aps,prb,reprint,showpacs,amsmath,amssymb,
floatfix,superscriptaddress,10pt]{revtex4-2}
\usepackage{graphicx}
\usepackage{float}
\usepackage{ulem}
\usepackage{mathrsfs}
\usepackage{braket}
\usepackage{color}
\usepackage{bbold}
\usepackage{bm}
\usepackage[mathlines]{lineno}
\usepackage[colorlinks=True,citecolor=blue,linkcolor=blue,
allcolors=blue,pdftoolbar   = false,
  pdfstartview={XYZ null null 1.25},bookmarks=false]{hyperref}
\usepackage{enumitem}
\usepackage{array, makecell}
\usepackage[utf8]{inputenc}
\usepackage{orcidlink}
\usepackage{tikz}
\newcommand*\circled[1]{\tikz[baseline=(char.base)]{
            \node[shape=circle,draw,inner sep=2pt] (char){#1};}}
\newcommand{\cosinv}{\cos^{-1}}

\def\red{\textcolor{red}}

\def\bea{\begin{eqnarray}}
\def\eea{\end{eqnarray}}
\def\bal{\begin{aligned}}
\def\eal{\end{aligned}}

\begin{document}
\title{Topologically protected dynamics in three-dimensional nonlinear antisymmetric Lotka-Volterra systems}
\author{Muhammad Umer\orcidlink{0000-0002-1941-1833}}
\email{umer@u.nus.edu}
\affiliation{Department of Physics, National University of Singapore, Singapore 117551, Republic of Singapore.}
\author{Jiangbin Gong\orcidlink{0000-0003-1280-6493}}
\email{phygj@nus.edu.sg}
\affiliation{Department of Physics, National University of Singapore, Singapore 117551, Republic of Singapore.}
\date{\today}

\begin{abstract}
Studies of topological bands and their associated low-dimensional boundary modes have largely focused on linear systems. This work reports robust dynamical features of three-dimensional ($3$D) nonlinear systems in connection with intriguing topological bands in $3$D. Specifically, for a $3$D setting of coupled rock-paper-scissors cycles governed by the antisymmetric Lotka-Volterra equation (ALVE)  that is inherently nonlinear, we unveil distinct characteristics and robustness of surface polarized masses and analyze them in connection with {the dynamics and} topological bands of the linearized Lotka-Volterra (LV) equation. Our analysis indicated that insights learned from Weyl semimetal phases with type-I and type-II Weyl singularities based on a linearized version of the ALVE are still remarkably useful, even though the system dynamics is far beyond the linear regime.   This work indicates the relevance and importance of the concept of topological boundary modes in analyzing high-dimensional nonlinear systems and hopes to  stimulate another wave of topological studies in nonlinear systems.  
\end{abstract}
\maketitle

\textit{Introduction.}--- The concept of topological robustness was introduced to condensed-matter physics in  1980's \cite{Klitzing1980,Thouless1982,Haldane1988}. Continued studies on topological matter have led to many important discoveries of new phases of matter \cite{Kane2005, *Bernevig2006, *Fu2007, Hasan2010, *Qi2011, Burkov2011, *Wan2011, *Zyuzin2012, *Hosur2013, *Huang2015, Bomantara2016, *Bomantara2016a, Benalcazar2017, *Schindler2018, *Li2018, Ghorashi2020}, including topological insulators \cite{Kane2005, *Bernevig2006, *Fu2007} and superconductors \cite{Hasan2010, *Qi2011, Tong2013},  Weyl semimetals \cite{Burkov2011, *Wan2011, *Zyuzin2012, *Hosur2013, *Huang2015, Bomantara2016, *Bomantara2016a}, as well as semimetals with linked and/or knotted nodal lines \cite{Li2018b}. Beyond the usual condensed-matter context,  topological phases of matter, especially their robust boundary states protected by topology,  have advanced studies of photonic crystals \cite{Haldane2008, *Raghu2008, *Wang2009}, acoustic systems \cite{Peng2016}, mechanical meta-materials \cite{Kariyado2015, *Suesstrunk2015, *Nash2015}, soft matter \cite{Delplace2017, *Zhou2018, *Pedro2019}, and biological \cite{Prodan2009, *Yamauchi2020} systems. Remarkably, though topological band theory is based on the linear Schr\"{o}dinger (or Schr\"{o}dinger-like) equation with translational invariance, it is also relevant to the understanding of robust boundary behavior in inherently nonlinear systems \cite{Knebel2020, Yoshida2021, Yoshida2021a}. This recognition is expanding the territory of topological physics with a potentially long-term impact. 

In this letter, we report how a well-established concept of topological gapless phases in $3$D systems, known as Weyl semimetal phases and their variants, can emerge in, and guide our understanding of a class of nonlinear systems, the so-called antisymmetric Lotka-Volterra equation (ALVE) system of coupled rock-paper-scissors (RPS) cycles in a $3$D configuration.

Specifically, polarized masses in $3$D ALVE systems are found to display distinct behaviors in different parameter regimes, with the time-averaged masses localized at the surface. Further, the propagation of masses is chiral at certain surfaces and surface polarized time-averaged masses are robust against the perturbation of system parameters. Such robust features hint a possible important role of topology concerning $3$D topological phases. To reveal the underlying topological physics, we  linearize the LV equation, which then resembles to the Schr\"{o}dinger equation of spinless particle on a $3$D lattice. Indeed, the resultant linearized equation of motion is analogous to that in symmetry class-$A$ of the ten-fold way topological classification in condensed-matter physics. Surprisingly, the qualitative dynamical behavior of the nonlinear system
is in parallel with that of the linearized LV equation, such as boundary propagation of probability density and localization of time-averaged probability density at certain surfaces of the $3$D network.  A topological band analysis hence becomes necessary and useful: there are rich gapless Weyl semimetal phases, including type-I Weyl semimetal, type-II Weyl semimetal, and hybrid Weyl semimetal with coexisting type-I and type-II Weyl nodes. As seen below, these various Weyl semimetal phases are critical to digest different dynamical behaviors in the nonlinear system. These findings hence establish a strong link between topological band theory and $3$D nonlinear systems.  This work also indicates that a number of distinct topological semimetal phases may be effectively realized in nonlinear systems, even though the dynamics is far beyond the linear regime.   

\begin{figure}[hbt!]
\centering
\includegraphics[clip, trim=0.2cm 0.0cm 0.0cm 0.8cm, width=1.00\linewidth, height=0.60\linewidth, angle=0]{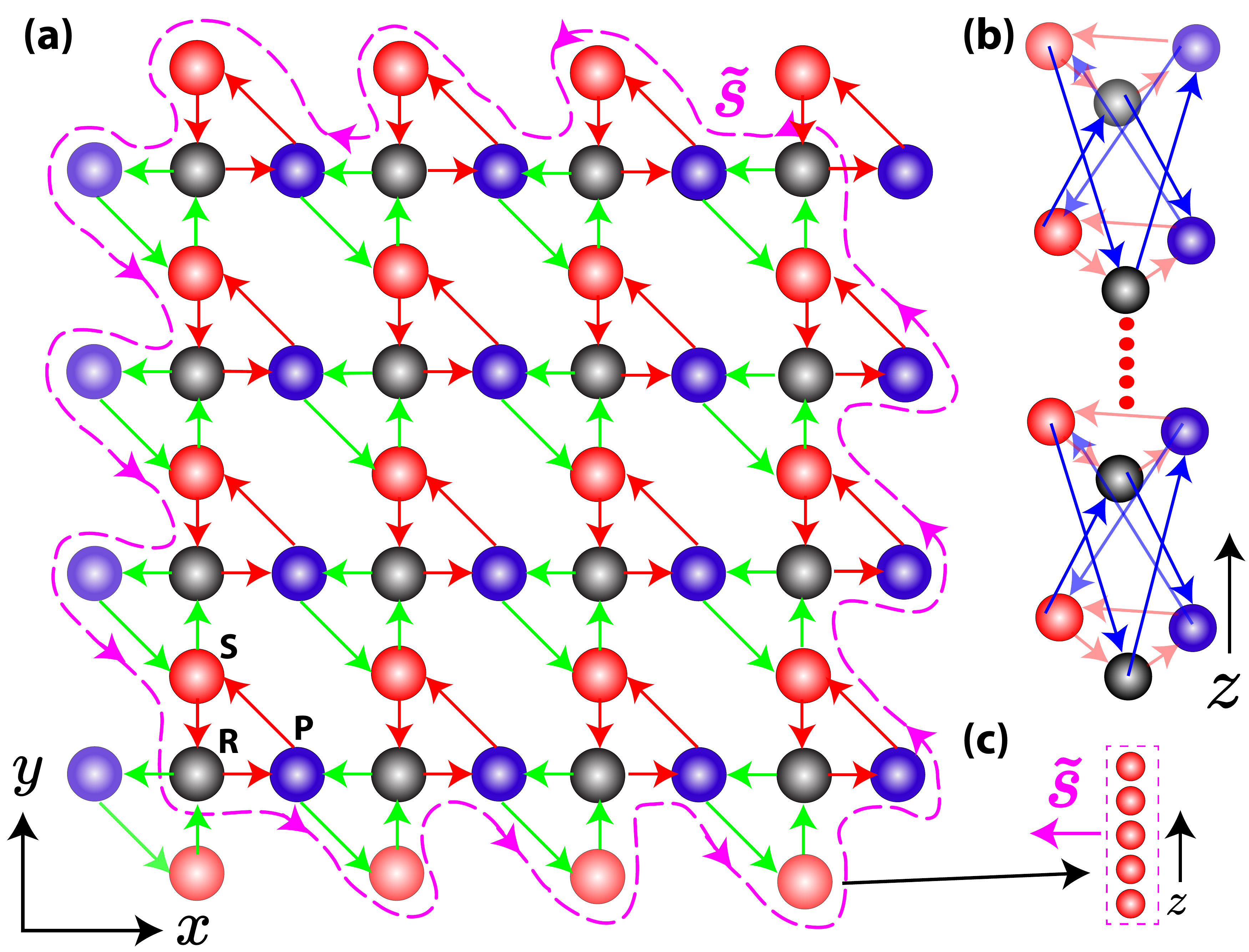}
\caption{Intra-cycle $t_{a}$ (red) and inter-cycle $t_{b},~t_{c}$ (green, blue) couplings in $x-y$ plane and $z$ direction are shown in panel (a, b). Magenta coloured path $\tilde{s}$ and arrows in panel (a) indicate the direction of chiral propagating masses an each layer while (c) shows that each point along the path $\tilde{s}$ has $N_{z}$ number of nodes stacked in the $z$ direction. Moreover, magenta path $\tilde{s}$ shows hard boundary for ALVE system while partially transparent nodes in (a, b) show system under periodic boundary conditions (PBC) in $x$, $y$, and $z$ direction. \vspace{-0.2cm}}
\label{Fig:Model}
\end{figure}

\textit{ALVE model system.}--- ALVE is a nonlinear model that describes the coexistence of species in game theory \cite{Knebel2013, *Knebel2015, *Toupo2015, *Geiger2018} and population dynamics \cite{Narendras1971}. 
The ALVE is given as
\bea\bal
\frac{\partial{u_{l}}}{\partial{t}} = u_{l}\sum^{M}_{m=1}\alpha_{lm}u_{m}\;,
\label{EQ:ALVE}
\eal\eea
where $\alpha_{lm}$\textquotesingle{s} are elements of the antisymmetric payoff matrix ${\bf A}$ with $\alpha_{ml} = -\alpha_{lm}$, $u_{l}~(u_{m})$ is the mass of the specie at site $l~(m)$, and $M$ is the total number of sites in the system. Mass on each site $u_{l}$ evolve under nonlinear interaction between masses $\sim u_{l}u_{m}$, where total mass remains conserved $\sum_{l = 1}^{M}\partial_{t}u_{l} = 0$. 

To explore the relevance of $3$D topological band theory in higher-dimensional nonlinear systems, we construct now a $3$D network system composed of stacked Kagome layers of coupled RPS game cycles. Each layer has $N$ number of sites arranged on the Kagome geometry in $x-y$ plane, as shown in Fig. \ref{Fig:Model}a. We then stack $N_{z}$ such layers in the $z$ direction for a $3$D network, as shown in Fig.~\ref{Fig:Model}b. Of more importance is the rules of the $3$D game. Within one or across different nearest neighbouring RPS cycles, the proposed rules are given as; R dominates S, S dominates P, and P dominates R, such that each strategy dominates (dominated by) only one other strategy \cite{Reichenbach2006, *Claussen2008, *Szolnoki2014}. These rules are implemented through intra-cycle $t_{a}$, inter-cycle payoff elements $t_{b}$ and $t_{c}$ in the $x-y$ plane and $z$ direction, respectively. The resulting ${\bf A} = \{\alpha_{lm}\}$ matrix is then found to be
\begin{small}
\bea\bal
&{\bf A} = \sum_{x = 1}^{N_{x}}\sum_{y = 1}^{N_{y}}\sum_{z = 1}^{N_{z}} \Big(t_{a}\big[\text{S}^{\dagger}_{x,y,z}\text{P}_{x,y,z} + \text{R}^{\dagger}_{x,y,z}\text{S}_{x,y,z}+ \text{P}^{\dagger}_{x,y,z}\text{R}_{x,y,z}\big] \\& + t_{b}\big[\text{P}^{\dagger}_{x,y,z}\text{R}_{x+1,y,z} - \text{S}^{\dagger}_{x,y,z}\text{R}_{x,y+1,z} + \text{S}^{\dagger}_{x,y,z}\text{P}_{x-1,y+1,z}\big]\\
&+ t_{c}\big[\text{P}^{\dagger}_{x,y,z}\text{R}_{x,y,z+1} + \text{P}^{\dagger}_{x,y,z+1}\text{R}_{x,y,z} + \text{R}^{\dagger}_{x,y,z}\text{S}_{x,y,z+1} \\ & + \text{R}^{\dagger}_{x,y,z+1}\text{S}_{x,y,z} + \text{S}^{\dagger}_{x,y,z}\text{P}_{x,y,z+1} + \text{S}^{\dagger}_{x,y,z+1}\text{P}_{x,y,z} \big] - h.c\Big),
\hspace{-0.4cm}\label{EQ:AEquation}
\eal\eea
\end{small}
\hspace{-0.15cm}where $\text{C}^{\dagger}_{x,y,z}$ ($\text{C}_{x,y,z}$) are creation (annihilation) bosonic operator for specie $\text{C} \in \{\text{R}, \text{P}, \text{S}\}$ at site $(x,~y,~z)$ of the $3$D network (See Supplemental Materials \footnote{see Supplemental Material at \red{insert link} for details} for matrix form) and $h.c$ is the hermitian conjugate. 

Here we assume that nodes are numbered first within a single layer, and then the same counting order continues to the second, third layer, etc. Furthermore, we consider strictly positive initial mass $u_{l}(t=0) = \sum_{j=1}^{N_{z}}\delta_{l,jN} + 1/M$ on each site $l$ such that mass is localized at $N^{\rm th}$-site (also at the hinge) of each layer in the system and rest of the nodes have background mass $\approx~1/M$. Then, ${\bf u}(t=0)$ can be regarded as a perturbed state away from the strictly positive steady-state (Nash equilibrium state) ${\bf c}$ that satisfies ${\bf Ac}= 0$, with $c_{l} = 1/M$. The dynamical features we explore below are in the vicinity of the system's steady-state. 

\begin{figure}[hbt!]
\centering
\includegraphics[clip, trim=0.0cm 0.6cm 0.0cm 2.8cm, width=1.00\linewidth, height=0.85\linewidth, angle=0]{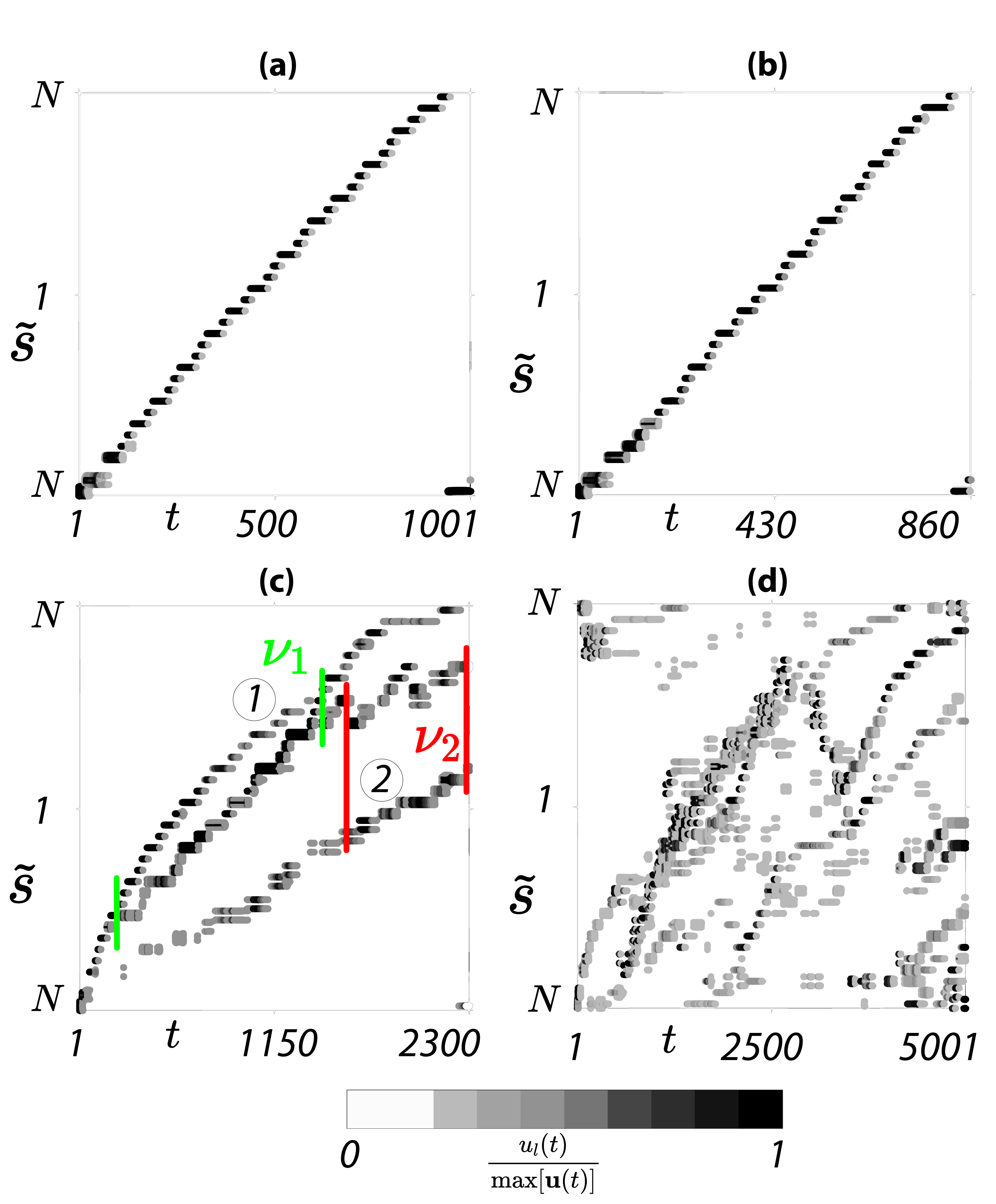}
\caption{Nonlinear evolution of masses on the surface $\tilde{s}$ of $3$D network are presented for parameter values (a) $t_{a} = 1.0$, $t_{c} = 3/8$, (b) $t_{a} = 1.4$, $t_{c} = 3/8$, (c) $t_{a} = 1.0$, $t_{c} = 5/8$, and (d) $t_{a} = 3.0$, $t_{c} = 4/10$, for fixed $t_{b} = 1.0$. \vspace{-0.2cm}}
\label{Fig:Dynamics}
\end{figure}

\begin{figure*}[hbt!]
\centering
\includegraphics[clip, trim=0.0cm 18.0cm 25.0cm 3.3cm, width=0.75\linewidth, height=0.38\linewidth, angle=0]{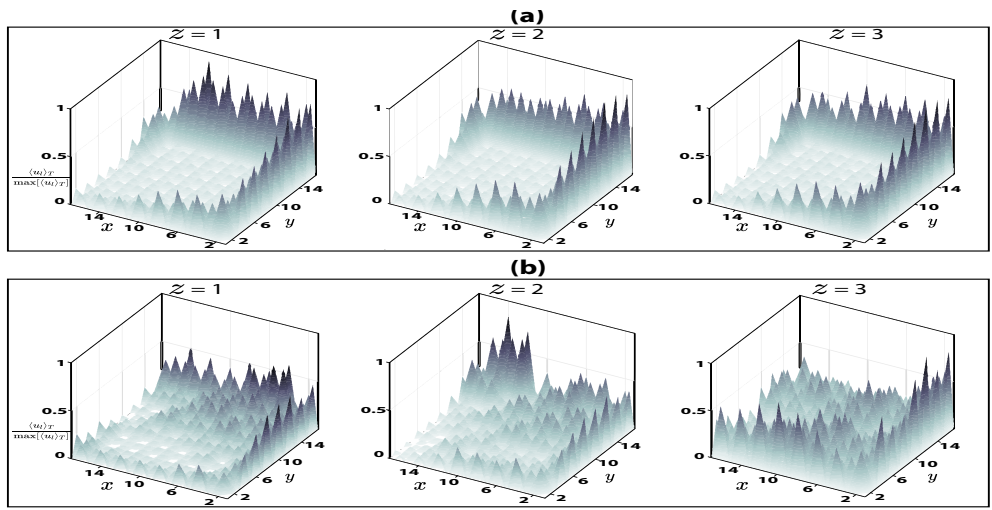}
\caption{Time-averaged profile of masses is shown for parameter values (a) $t_{a} = 1.0$, $t_{c} = 5/8$, $\Delta = 0.2$ and (b) $t_{a} = 3.0$, $t_{c} = 4/10$, for fixed $t_{b} = 1.0$, $N = 204$ and $N_{z} = 5$. Here $z$ indicate the layer in $z$-direction and we observe that for a system with $N_{z} = 5$, $z = 1~(2)$ and $z = 5~(4)$ have similarly localization of masses. \vspace{-0.2cm}}
\label{Fig:Localization}
\end{figure*}

\textit{Dynamical features in $3$D ALVE system.}--- The ALVE [Eq. (\ref{EQ:ALVE})] is computationally simulated and analysed, with $N = 52$ and $N_z=5$. First, we consider parameter values $t_{a} = 1.0$ and $t_{c} = 3/8$ (justification for this choice to be elaborated later). To better visualize the chiral time propagation of masses, we first define a path $\tilde{s}$ (see magenta coloured path in Fig. \ref{Fig:Model}a) that only contains the boundary nodes and passes through $1^{\rm st}$ and $N^{\rm th}$ nodes of each layer. At each position along the path $\tilde{s}$, there are $N_z$ different nodes due to their different coordinates along the $z$ direction. Fig. \ref{Fig:Dynamics}a depicts the time-evolving masses along the identified path $\tilde{s}$ (but also recording different behaviors at different layers), as a function of evolution time. Clearly, one sees that the movement of the masses is confined to the boundary of the system and can make a round trip (with an average velocity of $\nu = 0.0430$) if projected onto the $x-y$ plane. Another parameter selection, namely, $t_{a} = 1.4, ~t_{c} = 3/8$, correspond to analogous behavior of instantaneous masses $u_{l}(t)$ along the coordinate $\tilde{s}$ as shown in Fig. \ref{Fig:Dynamics}b, where average velocity ($\nu = 0.0470$) is relatively higher such that one complete round trip across the path $\tilde{s}$ takes less the time needed in Fig. \ref{Fig:Dynamics}a.

We now consider parameter values $t_{a} = 1.0, ~t_{c} = 5/8$. In this case, as shown in Fig.~ \ref{Fig:Dynamics}c, chiral propagation of masses splits into multiple waves moving at different velocities, a phenomenon  distinctively different from  the previously two situations. Specifically, the characteristic velocity of mass propagation of the middle wave in Fig.~\ref{Fig:Dynamics}c appears to assume two different values during two time windows. The upper wave $\circled{1}$ and middle wave in Fig.~\ref{Fig:Dynamics}c as bounded by the two green lines, propagate at a characteristic velocity of $\nu_{1} = 0.020$ node per unit time, whereas the lower wave $\circled{2}$ and middle wave as bounded by red lines, propagate at a characteristic velocity $\nu_{2} = 0.010$ node per unit time such that $\nu_{1}/\nu_{2} \approx 2$. Overall, it is observed that mass chiral propagation now has two characteristic velocities. To appreciate boundary localization, we consider the time-averaged behavior as some quasi-steady feature. In particular, we present the profile of time-averaged masses $\langle{u_{l}}\rangle_{T}$ in Fig. \ref{Fig:Localization}a, where boundary localization of time-averaged masses in $x-y$ plane can be observed, a feature shared by different layers.

It should also be highlighted that the above-observed localization behavior of the mass propagation is robust against the disorder among the system parameters. For example, disordered payoff matrix elements can be modelled by $\alpha_{lm} \pm \delta_{l}$ and $\alpha_{ml} \mp \delta_{l}$, where $\delta_{l} \in (-\Delta, \Delta)$ is derived from a uniform distribution of real numbers. In Fig.~\ref{Fig:Localization}a where  time-averaged surface polarized masses are presented, we have already averaged over one hundred realizations of disorder among the payoff matrix elements.

To conclude our investigations of different parameter regimes,  we now look into the dynamics for parameter values $t_{a} = 3.0, ~t_{c} = 4/10$. Fig. \ref{Fig:Dynamics}d indicates a totally different feature: the mass propagation is not chiral (without a clear velocity to identify) and the mass distribution is delocalized. Moreover, the time-averaged masses  are shown in Fig.~\ref{Fig:Localization}b for different layers. Delocalization of time-averaged masses across the entire $3$D system in Fig. \ref{Fig:Localization}b suggest the lack of a mechanism to confine the mass propagation at the surface of the system.

Above analysis implies that the $3$D ALVE system in different parameter regimes may be connected with distinct topological phases. Indeed, the observed robust boundary localization behavior in Fig.~\ref{Fig:Dynamics}a-\ref{Fig:Dynamics}c, and Fig. \ref{Fig:Localization}a resembles to the surface states protected by certain bulk topology in linear lattice models. If that is the case, then the absence of chiral mass propagation and boundary localization in Fig.~\ref{Fig:Dynamics}d and Fig. \ref{Fig:Localization}b suggests a topologically trivial regime.  To establish such connections, we linearize the ALVE system and carry out topological band theory analysis of the linearized system. 

\textit{Linearized LV equation and topological band analysis.}--- We reexpress the mass at node $l$ as $u_{l} = c_{l} + \delta{u_{l}}$, where $c_l$ is the steady-state mass and treat $\delta{u_{l}}$ as a small quantity. A linear approximation emerges once we take $\delta{u_{l}}\delta{u_{m}} \approx 0$, such that we have the following linearized LV equation \cite{Yoshida2021},
\bea\bal
i\partial_{t}{\bf \delta{u}} = \frac{i}{M} {\bf A} {\bf \delta{u}}\;.
\label{EQ:LV_Equation}
\eal\eea
It is noted that linearized LV equation becomes the Schr\"{o}dinger equation for ${\bf H} = i{\bf A}$ (up to a multiplicative factor of $1/M$, which only rescales the energy eigenvalues). Here the Hermitian matrix ${\bf H}$ can be taken as the Hamiltonian of a tight-binding system of spinless particle on a $3$D lattice.

Before we dive into the topological band theory analysis, let us examine the dynamical features of the linearized system described by Eq. (\ref{EQ:LV_Equation}). Interestingly, we observe similar qualitative features (see Supplemental Materials \cite{Note1} for numerical results) as in the nonlinear system described by Eq. (\ref{EQ:ALVE}), namely, surface propagation of probability density and surface localization of time-averaged probability density. The probability density wave propagate with some definite average velocity for parameter values similar to that of Fig. \ref{Fig:Dynamics}a and Fig. \ref{Fig:Dynamics}b. Moreover, probability density wave split into multiple waves propagating with two distinct velocities for the parameter values as given for Fig. \ref{Fig:Dynamics}c and these velocities have a ratio of $\approx 2$ \cite{Note1}. It is worth mentioning that the absolute velocities of probability density waves in the linearized systems here are different (quantitatively) as compared to that of surface propagating masses in nonlinear system. Such difference can be understood via a self-trapping mechanism typically induced by nonlinear interactions \cite{Ezawa2022} (note that the self-trapping mechanism may boost the chiral wave propagation velocity).  Interestingly, if we initiate the nonlinear dynamics by considering the starting configuration much closer to the steady-state, then the average velocity of mass propagation does approach that of the linear system (See Supplemental Materials \cite{Note1}). Finally, further suggesting the necessity of connecting the nonlinear dynamics with a linearized version, note that for parameter values  in Fig.~\ref{Fig:Dynamics}d, the probability density of the linear system does not propagate at the surface of the system and the time-averaged probability densities are also delocalized across the $3$D network.

We now introduce a Fourier transformation, under PBC in all spatial directions, from position to momentum space such that the bulk Hamiltonian $H({\bf k})$ in the momentum representation becomes
\bea\bal
&H({\bf k}) = t_{b}\sin(k_{x})\gamma_{1} - t_{b}\sin(k_{y})\gamma_{4} + t_{b}\sin(-k_{x} + k_{y})\gamma_{6} \\ &+ [t_{a} + t_{b}\cos(k_{x}) + 2t_{c}\cos(k_{z})]\gamma_{2} - [t_{a} + t_{b}\cos(k_{y})\\ & + 2t_{c}\cos(k_{z})]\gamma_{5} + [t_{a} + t_{b}\cos(-k_{x} + k_{y}) + 2t_{c}\cos(k_{z})]\gamma_{7}\;,
\hspace{-0.4cm}\label{EQ:Bulk_Ham}
\eal\eea
where $\gamma_{j}$\textquotesingle{s} are Gell-Mann matrices \cite{Note1}. 
The Hamiltonian Eq. (\ref{EQ:Bulk_Ham}) of linearized LV equation belongs to class-$A$ of ten-fold way symmetry classification of topological gapped systems \cite{Schnyder2008,*Ryu2010} and does not possess any internal symmetry such as charge conjugation, inversion, or time-reversal symmetry. 
Because symmetry class-$A$ has trivial classification for gapped system in $3$D,  only non-trivial gapless phases can emerge here.
 
\begin{figure}[hbt!]
\centering
\includegraphics[clip, trim=0.5cm 0.0cm 0.5cm 0.0cm, width=1.00\linewidth, height=1.00\linewidth, angle=0]{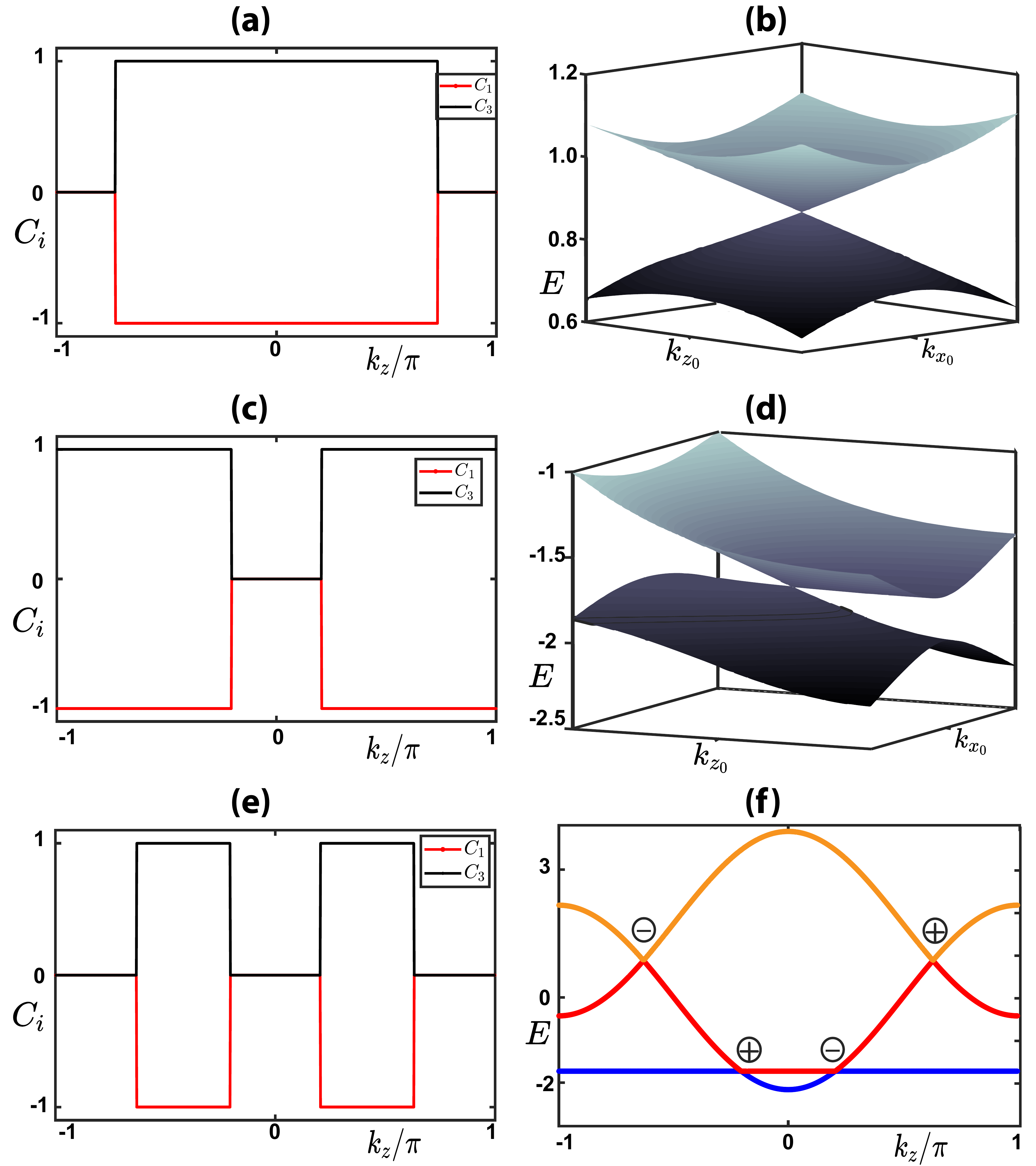}
\caption{Parameter values are taken as (a, b) $t_{a} = 1.0, t_{c} = 3/8$, (c, d) $t_{a} = 1.4, t_{c} = 3/8$, and (e, f) $t_{a} = 1.0, t_{c} = 5/8$, where (a, c, e) shows the slice Chern number $C(k_{z})$, (b, d) shows linear (tilted) dispersion of type-I (type-II) Weyl node for fixed $k_{y} = 2\pi/3$, and (f) shows coexisting type-I and type-II Weyl nodes for $(k_{x}, k_{y}) = (-2\pi/3, 2\pi/3)$. These type-I and type-II Weyl nodes have linear dispersion in $x-y$ direction such that $|\nu_{x}^{I}| = |\nu_{y}^{I}| = 0.0022$ and $|\nu_{x}^{II}| = |\nu_{y}^{II}| = 0.0044$ respectively, where we define $\nu^{I/II}_{i} = \partial{E}/\partial_{k_{i}}$. Moreover, it is noted that $\nu_{i}^{II}/ \nu_{i}^{I} \approx 2$ for $i = x~(y)$. \vspace{-0.2cm}}
\label{Fig:Phase}
\end{figure}

For fixed $t_{b} = 1.0$ without loss of generality, one can diagonalize the system for energy eigenvalues \cite{Note1} and it can be noted that two out of three bands touch at distinct momenta points given as $(k_{x_{0}}, k_{y_{0}}, k_{z_{0}}) = (2\pi/3, -2\pi/3, \pm\cosinv[\frac{\beta - t_{a}}{2t_{c}}])$, $(-2\pi/3, 2\pi/3, \pm\cosinv[\frac{\beta - t_{a}}{2t_{c}}])$ where inequality $|\frac{\beta - t_{a}}{2t_{c}}| < 1.0$ for $\beta \in \{t_{b}/2, 2t_{b}\}$ captures the band touching along $k_{z}$ momenta component. 
As we show in the following, these band touching points carry non-trivial topological charge. We thus refer to these points as Weyl nodes and corresponding phases as Weyl semimetals. 

First, we consider parameter values $t_{a} = 1.0$, $t_{c} = 3/8$ similar to that of Fig. \ref{Fig:Dynamics}a and inequality $|\frac{\beta - t_{a}}{2t_{c}}| < 1.0$ holds for $\beta = t_{b}/2$. The linearized LV system exhibit four type-I Weyl nodes, where Fig. \ref{Fig:Phase}b shows one of them for fixed $k_{y} = 2\pi/3$ with linear dispersion in $k_{x}-k_{z}$ momenta plane. Two Weyl nodes appear at two different energy values, for fixed $k_{z} = k_{z_0}$, where upper (lower) band touches the middle band \cite{Note1}. Charge and chirality of Weyl node located at (upper) lower energy can then be inferred from the (negative) change of Chern number \cite{Thouless1982} of (upper most) lowest band (-$C_{3}$) $C_{1}$ in Fig. \ref{Fig:Phase}a, while the Chern number of middle band $C_{2}(k_{z})$ is zero and does not change due to simultaneous touching with lower and upper bands at fixed $k_{z_{0}}$ (see \cite{Note1} for surface-state Fermi arc spectrum). 

Second, we consider parameter values $t_{a} = 1.4$, and $t_{c} = 3/8$ similar to that of Fig. \ref{Fig:Dynamics}b and inequality $|\frac{\beta - t_{a}}{2t_{c}}| < 1.0$ holds for $\beta = 2t_{b}$. The system exhibits a total of four type-II Weyl nodes where one of them is shown in Fig. \ref{Fig:Phase}d whose tilted nonlinear (flat) dispersion along $k_{z}$ can be observed for fixed $k_{y} = 2\pi/3$. As we discussed above, the Chern number of lower $C_{1}$ and negative of upper band $-C_{3}$ captures the charge and chirality of these Weyl nodes as we vary the momenta $k_{z}$ in Fig. \ref{Fig:Phase}c. 

Third, we consider parameter values $t_{a} = 1.0, t_{c} = 5/8$ similar to that of Fig. \ref{Fig:Dynamics}c and inequality $|\frac{\beta - t_{a}}{2t_{c}}| < 1.0$ now holds for $\beta = t_{b}/2$ and $\beta = 2t_{b}$ simultaneously. The linearized LV system has a total of eight Weyl nodes \cite{Note1} as compared to four in the previous two cases, out of which half (other half) of them are type-I (type-II). Linear and tilted nonlinear dispersion of these coexisting type-I and type-II Weyl nodes can be observed in Fig. \ref{Fig:Phase}f for fixed $(k_{x}, k_{y}) = (2\pi/3, -2\pi/3)$, where $\pm$ sign shows the chirality of corresponding Weyl node which can be duly inferred from Chern number $C(k_{z})$ in Fig. \ref{Fig:Phase}e. Moreover, one can obtain topological trivial phase for parameter values similar to that of Fig. \ref{Fig:Dynamics}d and Fig. \ref{Fig:Localization}b, where linearized LV system neither has any Weyl node nor surface states and each band has trivial topology \cite{Note1}.

\textit{Discussion.}--- We have shown that dynamical features of $3$D nonlinear ALVE system exhibit topological features of polarization and robustness against the disorder of system parameters. The underlying linearized LV equation also exhibits similar qualitative features of surface propagation and localization. Chiral propagation of mass (probability density) wave  can be associated with topological Weyl semimetal phases (of type-I or type-II) for the linearized LV equation. {On the other hand, chiral propagation of multiple mass waves in Fig.~\ref{Fig:Dynamics}c can be associated with a hybrid Weyl semimetal phase, where type-I and type-II Weyl nodes coexist in the linearized system.  Non-chiral and delocalized propagation in Fig.~\ref{Fig:Dynamics}d, \ref{Fig:Localization}b are explained via a topological trivial phase. These results clearly indicate the important relevance of topological band analysis for $3$D nonlinear systems. 

With RPS game cycles experimentally studied in social circles \cite{Semmann2003, Wang2014}, a $3$D construction of such platforms is necessary to uncover the implications of high-dimensional topological phases in nonlinear systems.   Above discussed linearized LV system may be also realized in electrical circuits \cite{Lee2018} due to the full control over circuit design.  As encouraged by the richness of semimetal phases found from $3$D nonlinear models here,  it is also possible to envision the relevance of higher-order topological phases in nonlinear systems. 

It is a pleasure to acknowledge helpful discussions with Raditya Weda Bomantara. 
J. Gong acknowledges support from Singapore NRF Grant No. NRF- NRFI2017-04 (WBS No. R-144-000-378-281). \\

\clearpage
\onecolumngrid
\begin{center}
\textbf{\large Supplementary Materials}\end{center}
\setcounter{equation}{0}
\setcounter{figure}{0}
\setcounter{table}{0}
\renewcommand{\theequation}{S\arabic{equation}}
\renewcommand{\thefigure}{S\arabic{figure}}
\renewcommand{\thetable}{S\arabic{table}}
\renewcommand{\cite}[1]{\citep{#1}}

This supplementary material consists of three section. In Sec. \ref{Sec:Sec1}, we present the antisymmetric ${\bf A}$ matrix and discuss the localization of time-averaged masses in nonlinear system for various parameter values.  In Sec. \ref{Sec:Sec2}, we present the numerical results for the dynamics of linearized system. Furthermore, we discuss the dynamics of nonlinear system under different initial state with varying perturbations and show that the dynamics of nonlinear system approaches that of linear system for smaller perturbations. In Sec. \ref{Sec:Sec3}, we discuss the topological band analysis, band spectrum and surface state Fermi arcs. 
\section{Antisymmetric Matrix}\label{Sec:Sec1}
First, we present a small block of antisymmetric payoff matrix {\bf A} from Eq. (2) of the main text, to better illustrate the payoff elements $t_{j}$ \textquotesingle{s} for $j \in \{a, b, c\}$. 
This small block in the matrix form is given as,
\setcounter{MaxMatrixCols}{20}
\bea
{\bf A} = \begin{pmatrix}
      0&-t_{a}&t_{a}& 0&-t_{b}&0& 0&0&t_{b}& 0&\dots& 0&-t_{c}&t_{c} & \dots\\
	  t_{a}&0&-t_{a}& 0&0&0& 0&0&0& -t_{b}&\dots& t_{c}&0&-t_{c}& \dots\\
      -t_{a}&t_{a}&0& 0&0&0& 0&0&0& 0&\dots& -t_{c}&t_{c}&0& \dots\\
      0&0&0& 0&-t_{a}&t_{a}& 0&0&0& 0&\dots& 0&0&0& \dots\\
	  t_{b}&0&0& t_{a}&0&-t_{a}& 0&0&0& 0&\dots& 0&0&0& \dots\\
      0&0&0& -t_{a}&t_{a}&0& 0&0&0& 0&\dots& 0&0&0& \dots\\
      0&0&0& 0&0&0& 0&-t_{a}&t_{a}& 0&\dots& 0&0&0& \dots\\
      0&0&0& 0&0&0& t_{a}&0&-t_{a}& 0&\dots& 0&0&0& \dots\\
      -t_{b}&0&0& 0&0&0& -t_{a}&t_{a}&0& 0&\dots& 0&0&0& \dots\\
      0&t_{b}&0& 0&0&0& 0&0&0& 0&\dots& 0&0&0& \dots\\
            \vdots&\vdots&\vdots& \vdots&\vdots&\vdots& \vdots&\vdots&\vdots& \vdots&\ddots& \vdots&\vdots&\vdots& \vdots \\
      0&-t_{c}&t_{c}& 0&0&0& 0&0&0& 0&\dots& 0&-t_{a}&t_{a}& \dots\\
      t_{c}&0&-t_{c}& 0&0&0& 0&0&0& 0&\dots& t_{a}&0&-t_{a}& \dots\\
      -t_{c}&t_{c}&0& 0&0&0& 0&0&0& 0&\dots& -t_{a}&t_{a}&0& \dots\\
      \vdots&\vdots&\vdots& \vdots&\vdots&\vdots& \vdots&\vdots&\vdots& \vdots&\vdots& \vdots&\vdots&\vdots& \ddots \\
    \end{pmatrix}\;,~~
{\bf u} = \begin{pmatrix}
R_{x,y,z}  \\
P_{x,y,z} \\
S_{x,y,z} \\
R_{x+1,y,z}  \\
P_{x+1,y,z} \\
S_{x+1,y,z} \\
R_{x,y+1,z}  \\
P_{x,y+1,z} \\
S_{x,y+1,z} \\
S_{x-1,y+1,z} \\
\vdots\\
R_{x,y,z+1}  \\
P_{x,y,z+1} \\
S_{x,y,z+1} \\
\vdots 
\end{pmatrix}\;,
~~~\label{Eq.AMatrix}
\eea
where ${\bf u}$ represents an arbitrary state vector. 


Second, we present the time-averaged behavior of masses for the parameter values $t_{a} = 1.0, t_{c} = 3/8$ and $t_{a} = 1.4, t_{c} = 3/8$ in Fig. \ref{Fig:App_Localization}a and Fig. \ref{Fig:App_Localization}b respectively. These two cases correspond to the surface propagation of masses in a single wave with some definite average velocity as shown in Fig. 2(a-b) of the main text. 
We present the profile of time averaged masses $\langle{u_{l}}\rangle_{T}$ in Fig. \ref{Fig:App_Localization}a-b, where boundary localization of time-averaged masses can be observed for various layers in the $z$ direction. Here the average is performed over time $t = T$, in which the surface propagating mass wave completes one round trip around the surface of the system. Moreover, the above mentioned results are obtained after considering one hundred realizations of disorder among the payoff matrix elements. Thus, this indicates that the surface propagation of masses is robustness against the disorder among the system parameter.

\begin{figure}[hbt!]
\centering 
\includegraphics[clip, trim=0.0cm 18.0cm 25.0cm 0.0cm, width=0.95\linewidth, height=0.60\linewidth, angle=0]{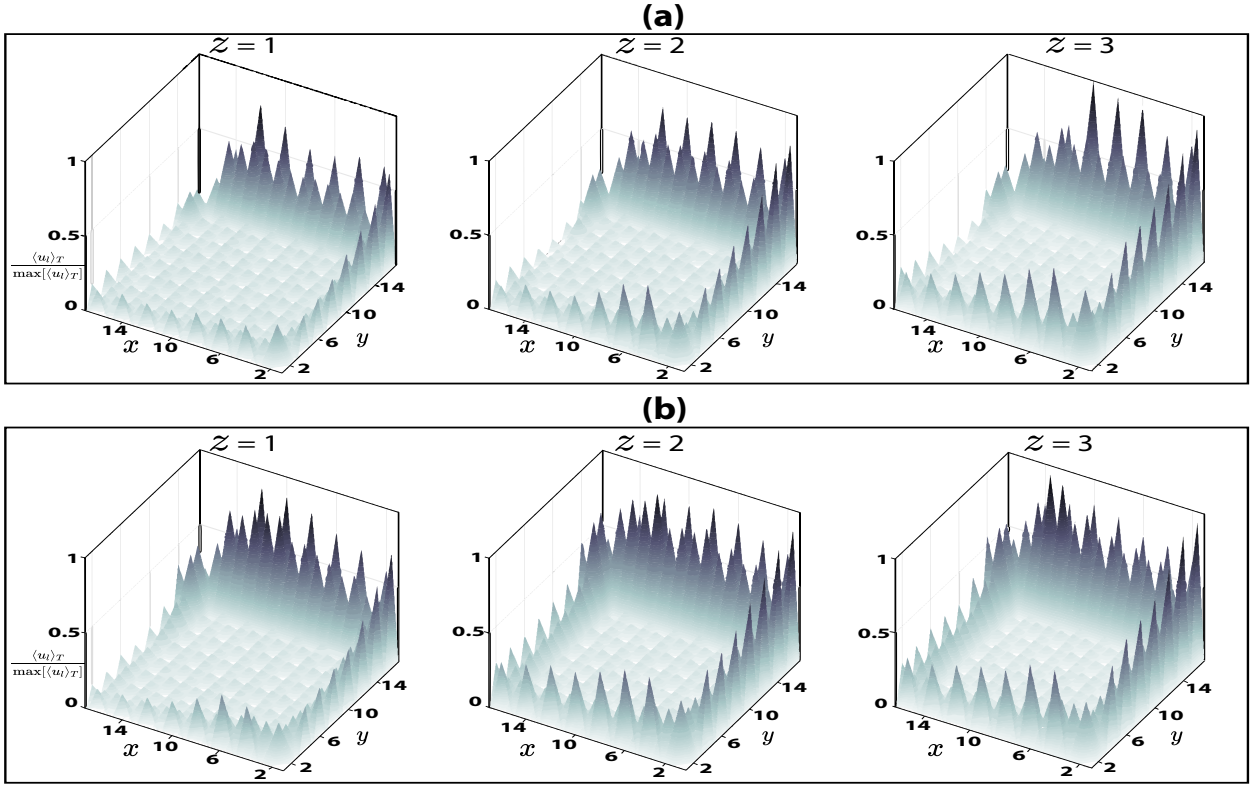}
\caption{Parameter values are considered as (a) $t_{a} = 1.0$, $t_{c} = 3/8$,  $\Delta = 0.2$, (b) $t_{a} = 1.4$, $t_{c} = 3/8$, $\Delta = 0.2$ for fixed $t_{b} = 1.0$,  $N = 204$ and $N_{z} = 5$. The averaging time $t = T$ is considered in which the surface propagating mass wave completes one round trip around the surface of the system. Layer $z = 4 ~(5)$ have similar localization as that of $z = 2 ~(1)$ respectively.}
\label{Fig:App_Localization}
\end{figure}

\section{\vspace{0.5cm}Linearized LV equation}\label{Sec:Sec2}
In order to construct the linearized LV equation, we follow the scheme discussed in Ref. \cite{Yoshida2021}. We consider a small perturbation in steady-state solution ($\partial_{t}{\bf c} = 0$ or $\partial_{t}c_{l} = 0$) of nonlinear system such that $u_{l} = c_{l} + \delta{u_{l}}$. Antisymmetric Lotka-Volterra equation then results in,
\bea \nonumber
\partial_{t}c_{l} + \partial_{t}\delta{u_{l}} &=& (c_{l} + \delta{u_{l}}) \sum_{m}\alpha_{lm}(c_{m} + \delta{u_{m}}),\\ \nonumber
\partial_{t}\delta{u_{l}} &=& c_{l}\sum_{m}\alpha_{lm}c_{m} + c_{l}\sum_{m}\alpha_{lm}\delta{u_{m}} + \delta{u_{l}}\sum_{m}\alpha_{lm}c_{m} + \delta{u_{l}}\sum_{m}\alpha_{lm}\delta{u_{m}} \\
\partial_{t}\delta{u_{l}} &=& c_{l}\sum_{m}\alpha_{lm}\delta{u_{m}},
\label{EQ:Linear}
\eea
where $\sum_{m}\alpha_{lm}c_{m} = 0$ and $\delta{u_{l}}\delta{u_{m}} \approx 0$ under linear approximation. We multiply Eq. (\ref{EQ:Linear}) by $i$ on both sides and obtain,
\bea
i\partial_{t}\delta{u_{l}} &=& \frac{i}{M}\sum_{m}\alpha_{lm}\delta{u_{m}},
\label{EQ:SE}
\eea
for $c_{l} = 1/M$. We recognize the Hamiltonian ${\bf H} = i{\bf A}$ upto a multiplicative factor of $1/M$ and $\delta{u_{l}}$ is the solution of this linear Schr\"{o}dinger (Schr\"{o}dinger-like) equation such that $|\delta{u_{l}}|^{2}$ describes the probability density. 

\begin{figure}[hbt!]
\centering 
\includegraphics[clip, trim=0.0cm 0.1cm 0.0cm 0.0cm, width=0.95\linewidth, height=0.60\linewidth, angle=0]{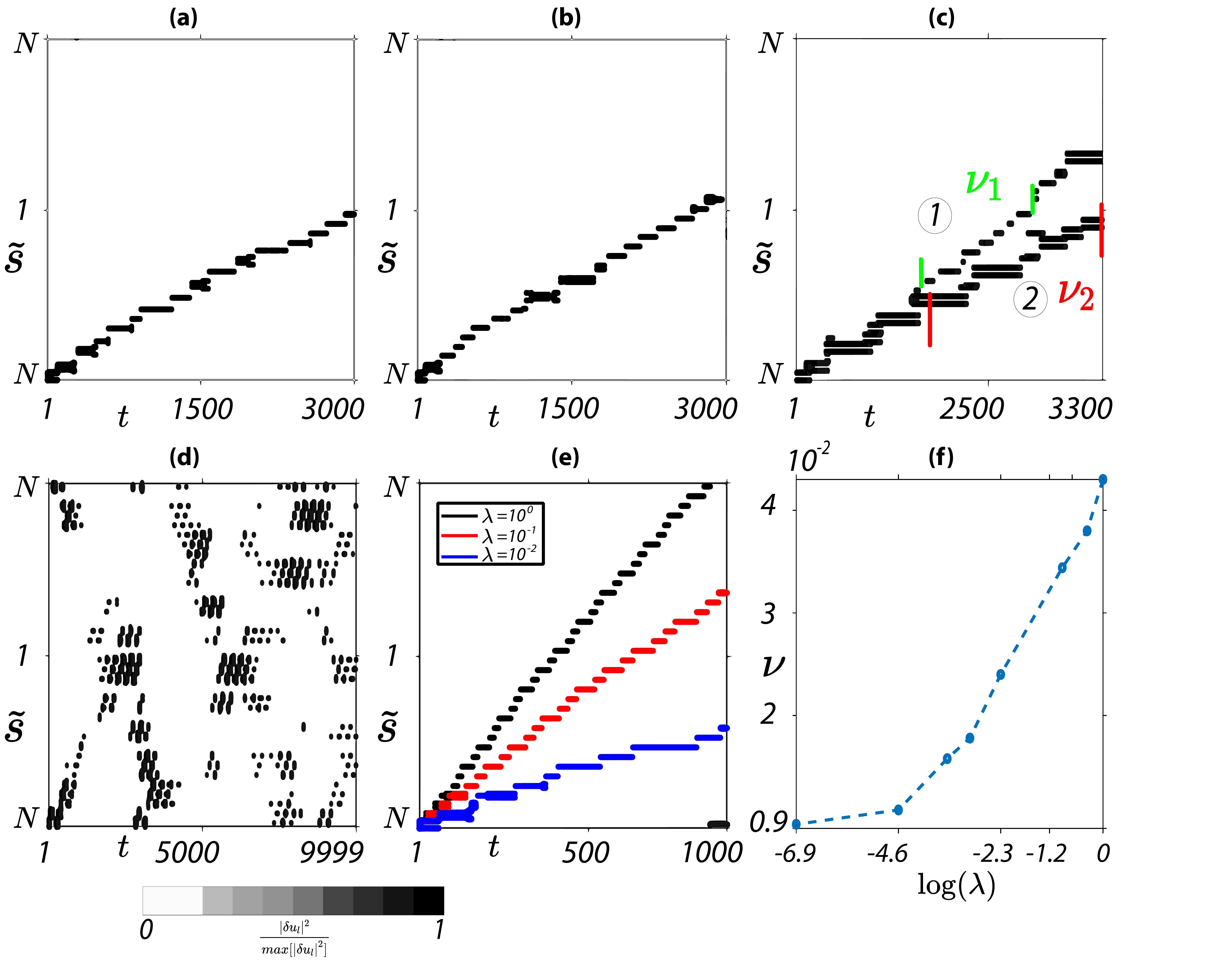}
\caption{Dynamics of initial state in linear system is presented for parameter values (a) $t_{a} = 1.0$, $t_{c} = 3/8$, (b) $t_{a} = 1.4$, $t_{c} = 3/8$, (c) $t_{a} = 1.0$, $t_{c} = 5/8$ and (d) $t_{a} = 3.0$, $t_{c} = 4/10$ for $N_{z} = 5, N = 52$ and $t_{b} = 1.0$. Panel (e,f) shows the dynamics and velocity of masses in nonlinear system as we vary the perturbation $\lambda$. Parameter values are taken as $t_{a} = 1.0, t_{c} = 3/8$ in (e,f). }
\label{Fig:App_Dynamics}
\end{figure}

We computationally solve this linear equation Eq. (\ref{EQ:SE}) for our system described by ${\bf H} = i{\bf A}$ where ${\bf A}$ is given by Eq. (2) of the main text. Before explaining the computational results, we address the question regarding the initial state of the linear system at $t = 0$. We consider the initial state of the system given by $\delta{\bf u} (t = 0) = \sum_{p}\mid\Psi_{\rm surface}^{p}\rangle\langle\Psi_{\rm surface}^{p}\mid{\bf u}\rangle$, where $u_{l} = \sum_{j = 1}^{N_{z}} \delta_{l,jN} + 1/M$ and $\mid \Psi_{\rm surface}^{p}\rangle$ is the $p^{th}$ surface state of the linear system ${\bf H}$. By construction, the initial state $\delta{\bf u} (t = 0)$ is localized at the surface of the $3$D system. It is important to highlight that the $\mid\Psi_{\rm surface}\rangle$ are either right or left moving surface states such that time evolution results in unidirectional propagation of state \cite{Ezawa2022}. Furthermore. it is worth noting that $\mid {\bf u}\rangle$ is similar to that considered for the nonlinear system in the main text. 

We now turn our attention towards the dynamics of the linear system whose results are presented in Fig. \ref{Fig:App_Dynamics}. To visualize the dynamics of the state at the surface of the system bounded by path $\tilde{s}$ (as shown in Fig. 1 of the main text), we present the probability density of the time evolving state in Fig. \ref{Fig:App_Dynamics}. It is observed that state evolves unidirectionally along the path $\tilde{s}$ at the surface of the system. For parameter values of $t_{a} = 1.0, t_{c} = 3/8$ and $t_{a} = 1.4, t_{c} = 3/8$, it is observed that the state evolves as a single wave of probability density as shown in Fig. \ref{Fig:App_Dynamics}a and Fig. \ref{Fig:App_Dynamics}b with an average velocity of $\nu \approx 0.0080$ and $\nu \approx 0.0090$ respectively. However, for parameter values $t_{a} = 1.0, t_{c} = 5/8$, the probability density wave split into two, which have two distinct propagation velocities $\nu_{1} \approx 0.0094$ and $\nu_{2} \approx 0.0052$ ($\nu_{1}/\nu_{2} \approx 2$) as shown in Fig. \ref{Fig:App_Dynamics}c. Finally, Fig. \ref{Fig:App_Dynamics}d shows the trivial scenario where probability density is delocalized and does not have chiral propagation. Qualitatively these dynamical features are the same as observed for the parent nonlinear system.  Quantitatively, however, the propagation velocities are different by roughly one order of magnitude.  This may be related to a self-trapping mechanism for chiral solitons \cite{Ezawa2022}.  

{To understand the above-mentioned magnitude difference in the wave propagation velocities between nonlinear and linear systems, we consider the nonlinear system with a varying initial state. We consider the initial state where mass on each site $l$ is given as $u_{l}(t = 0) = \sum^{N_{z}}_{j=1} {\lambda}\delta_{l,jN} + 1/M$, and $\lambda \le 1$ is a positive real number that tunes the magnitude of perturbation. By decreasing $\lambda$ from unity, one can move closer to the steady-state of the system (in the limit of $\lambda = 0$ one approaches the steady-state (Nash equilibrium state) ) of the nonlinear system.  In our system, $\lambda$ controls the interaction between $m^{th}$, and $(m-1)^{th}$ nodes which in turns affect the velocity of mass propagation. We consider various values of $\lambda$ and present our results in Fig. \ref{Fig:App_Dynamics}e-f. In Fig. \ref{Fig:App_Dynamics}e, we show that the masses propagate at the surface of $3$D nonlinear system with decreasing velocities, if we tune the initial state closer to the steady-state by decreasing the parameter $\lambda$. Compared to $\lambda = 1$, the velocity of mass propagation is much slower for $\lambda = 10^{-1}$ and $\lambda = 10^{-2}$. In fact, from Fig. \ref{Fig:App_Dynamics}f, one can observe that for $\lambda = 10^{-3}~ ({\rm log}(\lambda) \approx -6.9)$ the propagation velocity is $\nu \approx 0.009$, in agreement with the velocity of probability density $\nu = 0.0080$ in linear system for parameter values $t_{a} = 1.0, t_{c} = 3/8$. These results strengthen that main topological features of the linear system are highly useful to understand the dynamics at the system boundary, even when the dynamics is far from the linear regime.}

\section{\vspace{0.5cm}Energy spectrum and surface state Fermi arcs}\label{Sec:Sec3}
From equation ${\bf H} = i{\bf A}$, we consider periodic boundary conditions in all spatial directions followed by Fourier transformation from position to momentum space given as, 
\bea\bal
C_{x,y,z} = \frac{1}{\sqrt{N_{x}N_{y}N_{z}}}\sum_{k_{x},k_{y}, k_{z}}C_{k_{x},k_{y},k_{z}}e^{-i(k_{x} ax + k_{y}ay + k_{z}az)}\;,\\
C^{\dagger}_{x,y,z} = \frac{1}{\sqrt{N_{x}N_{y}N_{z}}}\sum_{k_{x},k_{y}, k_{z}}C^{\dagger}_{k_{x},k_{y},k_{z}}e^{i(k_{x}ax + k_{y}ay + k_{z}az)}\;,
\eal\eea
where $C \in \{\text{R}, \text{P}, \text{S}\}$, and lattice constant $a = 1$. Bulk Hamiltonian $H({\bf k})$ of the system is given by Eq. (4) of the main text such that ${\bf H} = \sum_{\bf k} \Psi^{\dagger}_{\bf k}H({\bf k})\Psi_{\bf k}$ and $\Psi^{\dagger}_{\bf k} = (\text{R}^{\dagger}_{\bf k}, \text{P}^{\dagger}_{\bf k}, \text{S}^{\dagger}_{\bf k})$. 
The traceless Gall-mann matrices in Eq. (4) of the main text are given as, 
\bea\nonumber
\gamma_{1} = \begin{pmatrix}
0 & 1 & 0 \\
1 & 0 & 0 \\
0 & 0 & 0 
\end{pmatrix}, ~~
\gamma_{2} = \begin{pmatrix}
0 & -i & 0 \\
i & 0 & 0 \\
0 & 0 & 0 
\end{pmatrix},~~
\gamma_{3} = \begin{pmatrix}
1 & 0 & 0 \\
0 & -1 & 0 \\
0 & 0 & 0 
\end{pmatrix},~
\gamma_{4} = \begin{pmatrix}
0 & 0 & 1 \\
0 & 0 & 0 \\
1 & 0 & 0 
\end{pmatrix},~~
\gamma_{5} = \begin{pmatrix}
0 & 0 & -i \\
0 & 0 & 0 \\
i & 0 & 0 
\end{pmatrix},~~
\eea
\bea\nonumber
\gamma_{6} = \begin{pmatrix}
0 & 0 & 0 \\
0 & 0 & 1 \\
0 & 1 & 0 
\end{pmatrix},~~
\gamma_{7} = \begin{pmatrix}
0 & 0 & 0 \\
0 & 0 & -i \\
0 & i & 0 
\end{pmatrix},~~
\gamma_{8} = \frac{1}{\sqrt{3}}\begin{pmatrix}
1 & 0 & 0 \\
0 & 1 & 0 \\
0 & 0 & -2 
\end{pmatrix}.
\eea

\begin{figure}[hbt!]
\centering 
\includegraphics[clip, trim=0.0cm 1.0cm 0.0cm 0.0cm, width=0.95\linewidth, height=0.50\linewidth, angle=0]{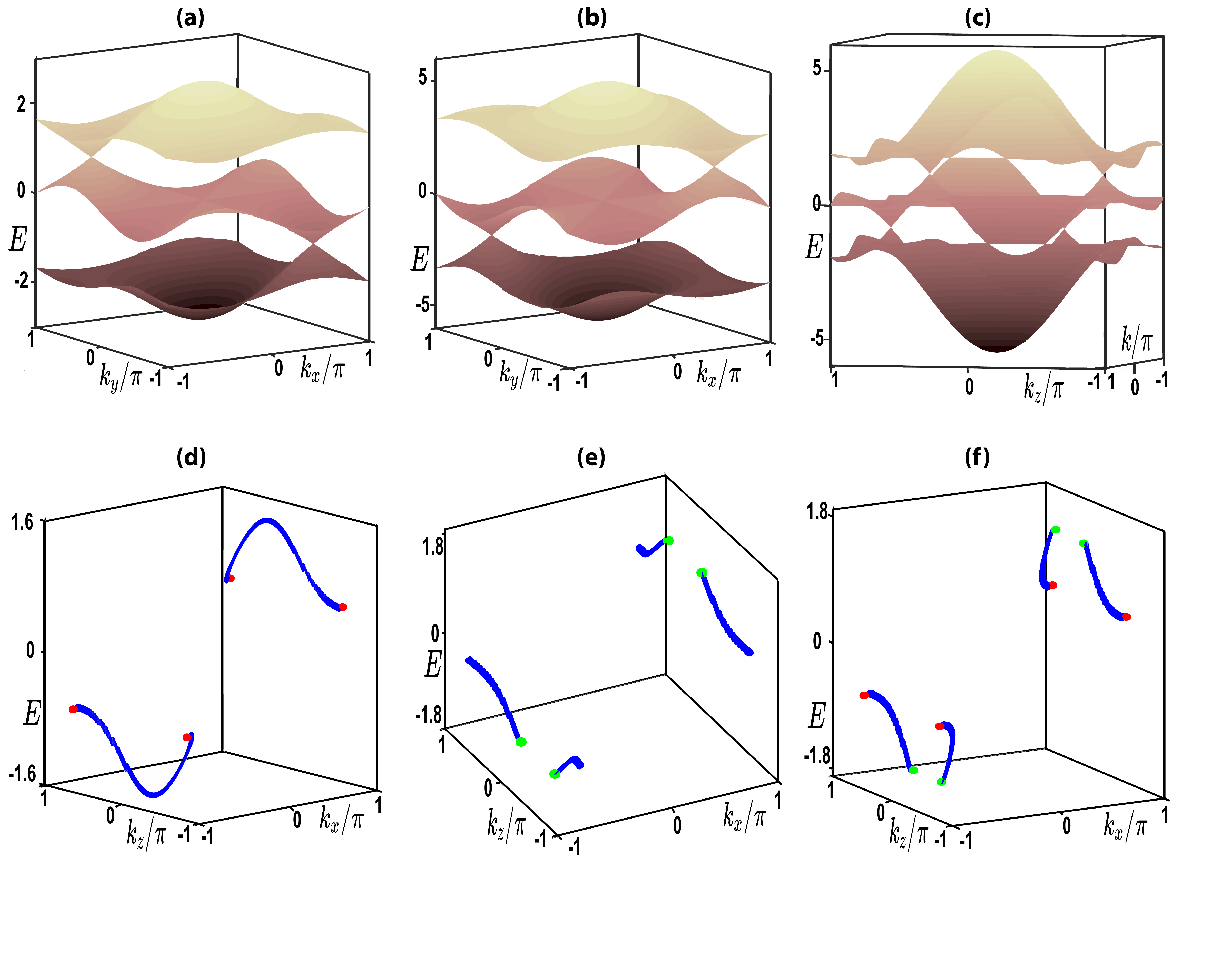}
\caption{Band structure and surface states Fermi arcs are shown for parameter values $t_{a} = 1.0, t_{c} = 3/8$ in (a, d), $t_{a} = 1.4, t_{c} = 3/8$ in (b, e), and $t_{a} = 1.0, t_{c} = 5/8$ in (c, f) for fixed $t_{b} = 1.0$. Red (Green) points shows the type-I (type-II) Weyl nodes while blue curve represent the surface state Fermi arcs.}
\label{Fig:App_Bands}
\end{figure}

We now diagonalize the bulk Hamiltonian for energy eigenvalues by solving the polynomial cubic equation $-E^{3} + E\theta_{1} + \theta_{2} = 0$ for fixed $t_{b} = 1$, whose solutions give the energy eigenvalues, 
\bea\bal\nonumber
E_{1} &= -\frac{2(3^{1/3})\theta_{2} + 2^{1/3}(-9\theta_{3} + \sqrt{ -12\theta_{2}^{3} + 81\theta_{3}^{2} }~)^{2/3}}{6^{2/3}(-9\theta_{3} + \sqrt{ -12\theta_{2}^{3} + 81\theta_{3}^{2} }~)^{1/3}}, \\ \nonumber
E_{2} &= \frac{(-1)^{1/3}\big( 2( 3^{1/3})\theta_{2} + (-2)^{1/3}(-9\theta_{3} + \sqrt{ -12\theta_{2}^{3} + 81\theta_{3}^{2} }~)^{2/3} \big)}{6^{2/3}(-9\theta_{3} + \sqrt{ -12\theta_{2}^{3} + 81\theta_{3}^{2} }~)^{1/3}}, \\
E_{3} &= \frac{ -2(-3)^{2/3}\theta_{2} + (-6)^{1/3}(-9\theta_{3} + \sqrt{ -12\theta_{2}^{3} + 81\theta_{3}^{2} }~)^{2/3} }{3(2^{2/3})(-9\theta_{3} + \sqrt{ -12\theta_{2}^{3} + 81\theta_{3}^{2} }~)^{1/3}},
\eal\eea
where $\theta_{1} = t_{a} + 2t_{c}\cos(k_{z})$, $\theta_{2} = 3 + 3\theta_{1}^{2} + 2\theta_{1}[\cos(k_{x}) + \cos(k_{y}) + \cos(k_{x} - k_{y})]$, and $\theta_{3} = 2(\theta_{1}^{2} -\theta_{1})[-\sin(k_{x}) + \sin(k_{y}) + \sin(k_{x}-k_{y})]$. One can solve for the relation $E_{i} = E_{j}$ ($i \neq j$) for the band touching points which results in Weyl nodes $(k_{x_{0}}, k_{y_{0}}, k_{z_{0}}) = (-2\pi/3, 2\pi/3, \pm\cosinv[\frac{\beta - t_{a}}{t_{c}}])$, and $(2\pi/3, -2\pi/3, \pm\cosinv[\frac{\beta - t_{a}}{t_{c}}])$ for $\beta = \{0.5, 2.0\}$. 
Fig. \ref{Fig:App_Bands}a, \ref{Fig:App_Bands}b, and \ref{Fig:App_Bands}c, show the band structure where Weyl nodes can be observed at $E_{1} = E_{2}$ and $E_{2} = E_{3}$ for fixed $k_{z} = k_{z_{0}}$. Type-I and type-II nature of these Weyl nodes has already been shown in Fig. 4 of the main text. 

Surface state Fermi arcs are shown in Fig. \ref{Fig:App_Bands}d, \ref{Fig:App_Bands}e, and \ref{Fig:App_Bands}f for three representative values of parameters under periodic boundary conditions in $x-z$ directions and open boundary conditions in $y$ direction where red (green) points indicate the type-I (type-II) Weyl nodes. Here, blue curve represent the points in energy-momenta space where $E_{\rm left} = E_{\rm right}$, with $E_{\rm left} (E_{\rm right})$ indicating the energy of left (right) moving surface states as a function of $k_{x}-k_{z}$ momenta under open boundary conditions in $y$ direction. It is worth noting that in hybrid Weyl semimetal phase, surface state Fermi arc connects a type-I Weyl node to a type-II Weyl node of opposite chirality. 


Finally, we note that all the Chern numbers for parameter values $t_{a} = 3.0$, $t_{b} = 1$, and $t_{c} = 4/10$ that correspond to Fig. 2d of the main text are found to be zero as we sweep through $k_{z}$. Thus, there are no Weyl nodes or surface states Fermi arcs in this case.

\bibliographystyle{apsrev4-2}
%

\end{document}